\documentclass[amsmath, amssymb,10pt,aps,prb,twocolumn,notitlepage,showpacs,superscriptaddress]{revtex4-2}
\usepackage{graphicx}
\usepackage{calc}
\usepackage{braket}
\usepackage{ulem}   
\usepackage{slashed}
\usepackage{wasysym}
\usepackage{siunitx}
\usepackage{dsfont}
\usepackage{amsthm,amsmath,amsfonts,amssymb,verbatim,color}
\usepackage{graphicx}
\usepackage{subfigure}
\usepackage{bm}
\usepackage{epsfig,slashed}
\usepackage[T1]{fontenc}
\usepackage[colorlinks=true,citecolor=blue,linkcolor=blue,urlcolor=blue]{hyperref}
\usepackage[version=4]{mhchem}
\usepackage{gensymb}
\usepackage{ulem}
\begin{document}
\title{Persistent Fluctuating Superconductivity and Planckian Dissipation in Fe(Te,Se)}

\author{Jonathan Stensberg}
\email{js6750@columbia.edu}
\affiliation{Department of Physics and Astronomy, University of Pennsylvania, Philadelphia, Pennsylvania 19104, USA}

\author{Pok Man Tam}
\affiliation{Department of Physics and Astronomy, University of Pennsylvania, Philadelphia, Pennsylvania 19104, USA}
\affiliation{Princeton Center for Theoretical Science, Princeton University,
Princeton, New Jersey 08544, USA}

\author{Xiaoyu Yuan}
\affiliation{Department of Physics and Astronomy, Rutgers, The State University of New Jersey, Piscataway, New Jersey 08854, USA}
\author{Xiong Yao}
\affiliation{Department of Physics and Astronomy, Rutgers, The State University of New Jersey, Piscataway, New Jersey 08854, USA}
\author{Heshan Yu}
\affiliation{Department of Materials Science and Engineering, University of Maryland, College Park, MD 20742, USA.}
\author{Chih-Yu Lee}
\affiliation{Department of Materials Science and Engineering, University of Maryland, College Park, MD 20742, USA.}

 \author{An-Hsi Chen}
    \affiliation{Materials Science and Technology Division, Oak Ridge National Laboratory, Oak Ridge, Tennessee 37831, USA}

 \author{Philip J.D. Crowley}
    \affiliation{Department of Physics, Harvard University, Cambridge, MA, 02138, USA}
    \affiliation{Department of Physics and Astronomy, Michigan State University, East Lansing, Michigan 48824, USA}

 \author{Matthew Brahlek}
    \affiliation{Materials Science and Technology Division, Oak Ridge National Laboratory, Oak Ridge, Tennessee 37831, USA}
    
\author{Ichiro Takeuchi}
\affiliation{Department of Materials Science and Engineering, University of Maryland, College Park, MD 20742, USA.}
\affiliation{Maryland Quantum Materials Center, Department of Physics, University of Maryland, College Park, Maryland 20742, USA}
    
\author{Seongshik Oh}
\affiliation{Department of Physics and Astronomy, Rutgers, The State University of New Jersey, Piscataway, New Jersey 08854, USA}

\author{Joseph Orenstein}
\affiliation{Department of Physics, University of California, Berkeley, California 94720, USA}
\affiliation{Materials Science Division, Lawrence Berkeley National Laboratory, Berkeley, California 94720, USA}

\author{Charles Kane}
\affiliation{Department of Physics and Astronomy, University of Pennsylvania, Philadelphia, Pennsylvania 19104, USA}

\author{Liang Wu}
\email{liangwu@sas.upenn.edu}
\affiliation{Department of Physics and Astronomy, University of Pennsylvania, Philadelphia, Pennsylvania 19104, USA}

\date{\today}

\begin{abstract}
Increasingly intricate phase diagrams in new classes of superconductors host fascinating interactions between superconductivity, diverse quantum phases, and quantum critical dynamics. The native superfluids, however, often exhibit much lower density and much greater inhomogeneity than conventional superfluids. This may render the superconductivity susceptible to fluctuations that are ordinarily assumed to be frozen out far below the superconducting transition temperature $T_c$, calling into question the degree to which the superconducting state is fully coherent. In this work, we leverage terahertz spectroscopy to demonstrate strongly fluctuating superconductivity in topological compositions of the multiband iron-based superconductor Fe(Te,Se). These fluctuations are found to persist undiminished far below $T_c$ and converge upon the limit of Planckian dissipation above $T_c$. These results indicate that extended quantum fluctuations dominate the electrodynamics of both the superconducting and Planckian-dissipative precursor states of Fe(Te,Se), and demonstrate that the assumption of phase coherence must be rigorously validated in emerging classes of unconventional superconductors.

\end{abstract}

\pacs{}
\maketitle

\section{Introduction}\label{sec1}

The iron-chalcogenides host a rich ensemble of unconventional superconducting states emerging from distinct electron and hole bands \cite{Kreisel2020,Fernandes2022}. The Fe(Te,Se) system presents a particularly fruitful playground with multiple layers of interacting physics, including strong-coupling superconductivity nearing the Bardeen-Cooper-Schrieffer - Bose-Einstein condensate (BCS-BEC) crossover regime \cite{Lubashevsky2012}, time-reversal symmetry breaking that emerges alongside the superconducting state \cite{Zaki2021}, coupling to electronic nematicity that suppresses superconductivity \cite{Zhao2021}, and topological superconductivity \cite{Zhang2018} hosting purported Majorana physics \cite{Wang2018,Zhu2020} useful for topological quantum computing. Yet despite these many interactions and underappreciated studies revealing quenched disorder \cite{Machida2019}, spatially-varying phases \cite{Li2021}, and superfluid inhomogeneity \cite{Cho2019}, the robust coherence of the superconducting state in Fe(Te,Se) is generally taken for granted.

To the contrary, we demonstrate the persistence of fluctuating superconductivity throughout the entire superconducting regime of Fe(Te,Se) using time-domain terahertz spectroscopy (TDTS). Strong fluctuations persist down to the lowest temperatures of $\sim$ 20$\%$ of the superconducting transition temperature $T_c$, converge upon the limit of Planckian dissipation at $T_c$, and persist into the precursor state above $T_c$. Analysis of these fluctuations indicates a complex interplay between coexistent weak-coupling superconductivity emerging from the hole band, strong-coupling superconductivity emerging from the electron band, and the Planckian-dissipative state, often associated with quantum criticality \cite{HartnollMackenziePRMP2022,PhillipsScience2022}. These behaviors are evidenced by a comprehensive set of anomalous electrodynamics, establish strong fluctuations and Planckian dissipation as key features of the superconducting and precursor states in Fe(Te,Se), and further highlight the role of both fluctuations and Planckian dissipation across multiple classes of superconductors.

\section{Anomalous Terahertz Electrodynamics}\label{sec_THz}

\begin{figure*}
    \centering
    \includegraphics[width=\textwidth]{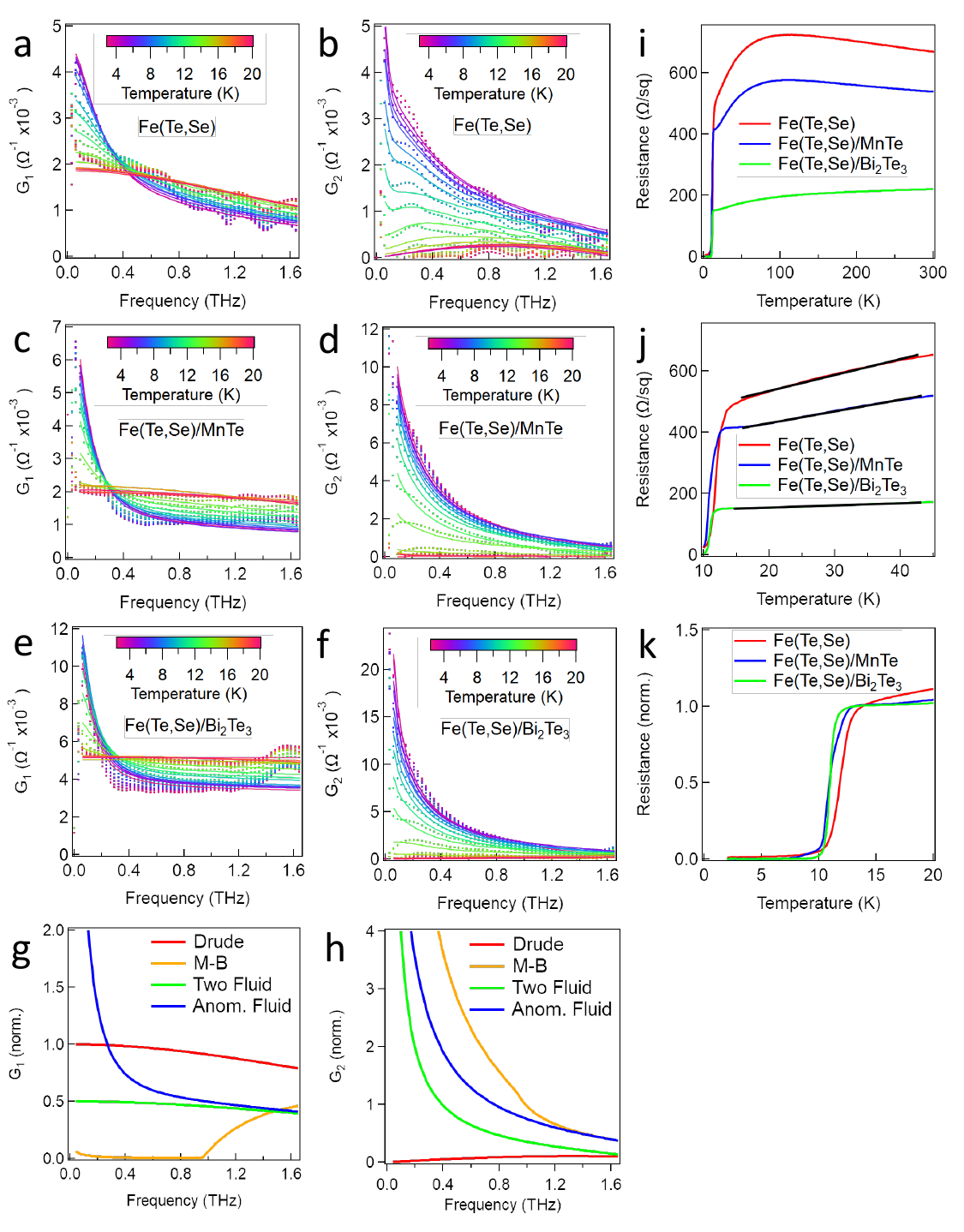}
    \caption{\textbf{Terahertz electrodynamics of Fe(Te,Se).} Real and complex terahertz conductance of \textbf{(a, b)} Fe(Te,Se) thin film, \textbf{(d, d)} Fe(Te,Se)/MnTe heterostructure, and \textbf{(e, f)} Fe(Te,Se)/Bi$_2$Te$_3$ heterostructures, respectively. Solid lines show the fitting of the anomalous fluid model to the data points for each temperature. \textbf{(g, h)} Sketches of real and imaginary, respectively, for the optical conductivity of Drude, Mattis-Bardeen, Two-Fluid, and Anomalous Fluid models. \textbf{(i, j, k)} Sample resistance measured in DC transport showing full temperature dependence, linear-in-$T$ behavior over 15-45 K, and broad multi-Kelvin superconducting transition, respectively.}
    \label{fig:Fig1}
\end{figure*}

We probe the electrodynamics across the superconducting transition of various Fe(Te,Se) samples via TDTS. This technique achieves coherent measurement of the complex conductance $\Tilde{G} = G_1 + iG_2$, which encodes the low-energy electrodynamics (1 THz $\sim$ 4.1 meV) of superconductors with high precision \cite{Stensberg2023}. (See Methods for TDTS details). The TDTS data for three primary Fe(Te,Se) samples is shown in Fig \ref{fig:Fig1}.a-f spanning the superconducting transition, with corresponding DC transport data in Fig \ref{fig:Fig1}.i-j. Data for additional samples can be found in the Supplementary Information (SI). The three primary samples are thin films of FeTe$_{0.7}$Se$_{0.3}$ grown by molecular beam epitaxy to a thickness of $t = 20$ nm. One sample is grown directly on a CaF$_2$ substrate, and two heterostructures are grown via hybrid symmetry epitaxy \cite{Yao2021,Yao2022} atop 10 nm of antiferromagnetic MnTe or topological insulator Bi$_2$Te$_3$ (See Methods for extended growth details). By changing the material upon which the Fe(Te,Se) is grown, slight sample-to-sample variations are introduced while preserving key properties of Fe(Te,Se). In particular, all three samples exhibit the linear-in-$T$ resistance characteristic of a strange metal \cite{PhillipsScience2022,HartnollMackenziePRMP2022} from $\sim$15 K to 40$+$ K, and all three show a transition to superconductivity that begins around 13 K and spans a few Kelvin, highlighted respectively in Fig. \ref{fig:Fig1}.j,k.  

At 20 K, all three samples show similar electrodynamic behaviors in Fig. \ref{fig:Fig1}.a-f: a finite $G_1$ and near-zero $G_2$ consistent with a Drude-like conductance of the Hund's metal ground state of Fe(Te,Se) above $T_c$ \cite{Kreisel2020,Fernandes2022}. This behavior persists down to about 13 K, the temperature at which the superconducting transition first appears in the DC transport data. Given this concurrence, we will use $T_c \approx 13$ K throughout unless otherwise noted. As the superconducting transition unfolds, the terahertz electrodynamics steadily evolve with decreasing temperature, ultimately converging upon a remarkable low-temperature electrodynamic profile with three key features: (1) a roughly 1/$\omega$-like $G_2$, (2) a very sharp low-frequency peak in $G_1$, and (3) a depreciated yet significant high-frequency background in $G_1$.

This distinctive behavior is not described by either of the standard models for the dynamical conductivity of superconductors, namely the Mattis-Bardeen \cite{MattisBardeen1958} and two-fluid models \cite{BasovTimusk2005}. As sketched in Fig. \ref{fig:Fig1}.g,h, both models reproduce feature (1), namely the hallmark $G_2 \propto 1/\omega$ behavior of a superconductor, but $G_1$ disagrees greatly. The Mattis-Bardeen model for conventional BCS-like superconductors features a stark suppression of $G_1$ at low frequencies below the pair-breaking energy of twice the superconducting gap $2\Delta$. In contrast, the two-fluid model, most commonly used to describe cuprate superconductors, has a broad but shallow depreciation of $G_1$, corresponding to a large residual normal fluid due to incomplete condensation of Cooper pairs. The two-fluid model addresses feature (3), namely the large residual $G_1$ seen in Fig. \ref{fig:Fig1}.a,c,e; however, it is ultimately inadequate to simultaneously describe feature (2), the sharp low-frequency peak in $G_1$. Introducing an additional Drude-like fluid is necessary to address this low-frequency peak in $G_1$ and fully capture features (1)-(3). The origin of this added anomalous fluid contribution will be examined at length below. This yields the following model for the electrodynamics of Fe(Te,Se) across the superconducting transition and sketched as the anomalous fluid model in Fig. \ref{fig:Fig1}.g,h:
\begin{equation}   
    G_1 = k_n\frac{\tau_n}{(1+\tau_n^2 \omega^2)} + k_s\delta(\omega) + k_a\frac{\tau_a}{(1+\tau_a^2 \omega^2)} 
\end{equation}
\begin{equation}
    G_2 = k_n\frac{\tau_n^2\omega}{(1+\tau_n^2 \omega^2)} + k_s\frac{1}{\omega} + k_a\frac{\tau_a^2\omega}{(1+\tau_a^2 \omega^2)} + (\epsilon_\infty - 1) \omega
\end{equation}
\noindent where the $k_i$ give the weights of the normal fluid $k_n = n_ne^2t^*/m_n^*$, superfluid $k_s = n_se^2t^*/m_s^*$, and anomalous fluid $k_a = n_ae^2t^*/m_a^*$ ($e$ = electron charge, $m_i^*$ = effective mass, $t^*$ = effective sample thickness, $n_i$ = carrier density). The final term accounts for lattice polarizability $\epsilon_\infty$ from absorption above the measured range. As shown by the solid lines in Fig. \ref{fig:Fig1}.a-f, the anomalous fluid model provides a strong fit and accounts for all three key features of the terahertz electrodynamics of Fe(Te,Se).

\begin{figure*}
    \centering
    \includegraphics[width=\textwidth]{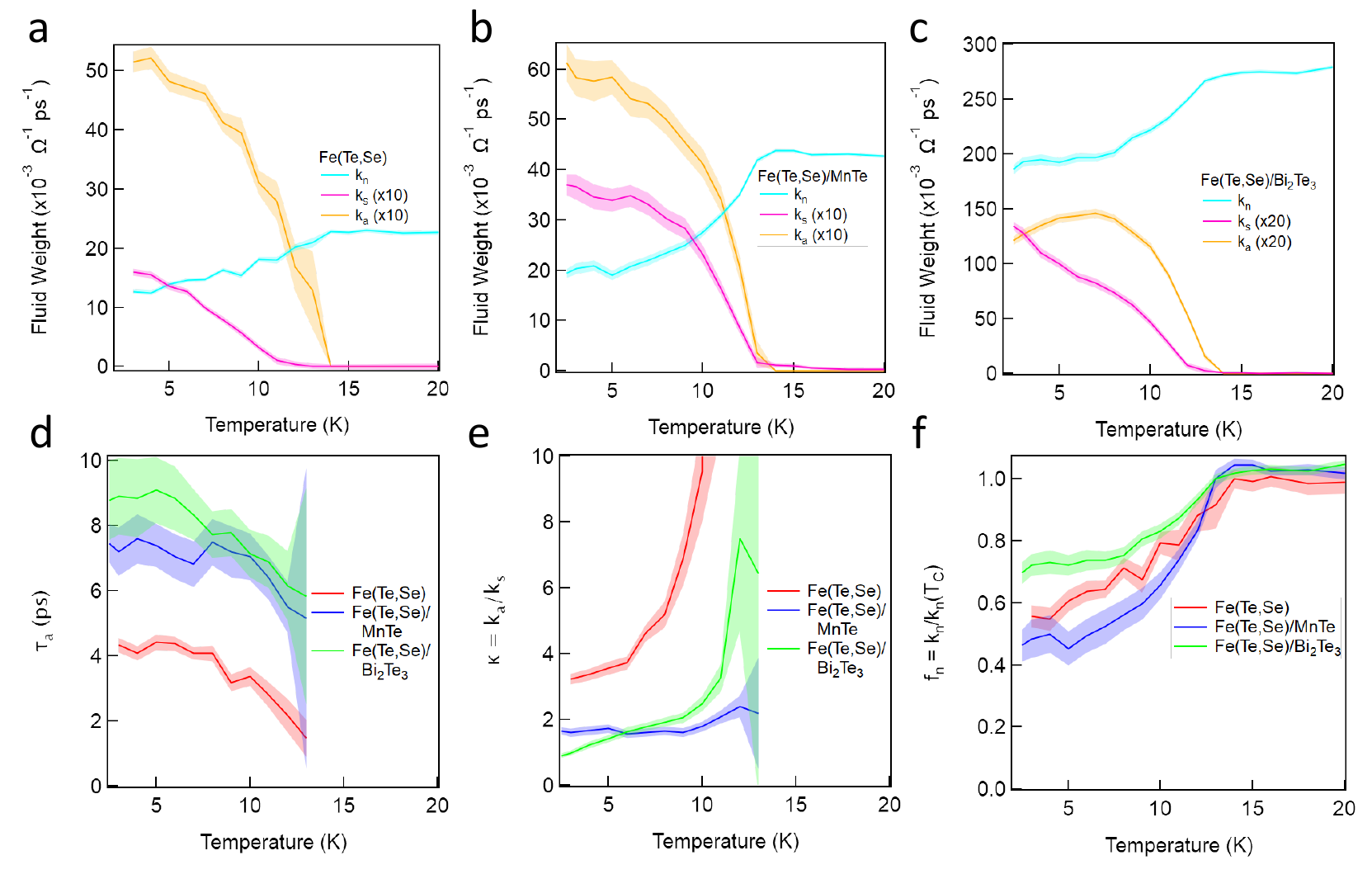}
    \caption{\textbf{ Anomalous fluid model of Fe(Te,Se).} \textbf{a-c} Fitted values of the normal fluid, superfluid, and anomalous fluid weights of Fe(Te,Se), Fe(Te,Se)/MnTe, and Fe(Te,Se)/Bi$_2$Se$_3$ samples, respectively. \textbf{d.} Fitted values of scattering time for the anomalous fluids. \textbf{e.} Proportionality $\kappa=k_a/k_s$ of the anomalous fluid weight to the superfluid weight. \textbf{f.} Fraction of normal fluid at each temperature. Uncertainties in fitted values are given by the shaded region.}
    \label{fig:Fig2}
\end{figure*}

The key results of fitting the anomalous fluid model are shown in Fig. \ref{fig:Fig2}. Above $T_c$, the data is fully described by a Drude function with scattering times of $\tau_n \lesssim 0.1$ ps. Below $T_c$, both the superfluid and anomalous fluid emerge simultaneously with the depreciation of the normal fluid, as shown by the fitted fluid weights in Fig. \ref{fig:Fig2}.a-c.  As indicated by Fig. \ref{fig:Fig2}.d, the anomalous fluid features a sub-terahertz scattering rate whose mild temperature dependence appears to track the superfluid weight. These features suggest the superfluid and the anomalous fluid are fundamentally related, with their weights appearing to be roughly proportional to the eye. Quantitatively, the ratio of the two fluid weights, $\kappa = k_a/k_s$, shown in Fig. \ref{fig:Fig2}.e shows a fairly stable proportionality at low temperature that sharply peaks just below $T_c$. These behaviors coincide with a large residual normal fluid fraction, shown by $f_n = k_n(T)/k_n(T_c)$ in Fig. \ref{fig:Fig2}.f, indicating the stable coexistence of condensed Cooper pairs and uncondensed normal carriers even at the lowest temperatures $T < 0.2T_c$. Altogether, the terahertz electrodynamics indicate a fascinating picture: the stable coexistence even in the low-temperature limit of a seemingly-lossy superfluid, generating features (1) and (2), with a residual normal fluid that appears unaware of this superfluid apart from the reduction in its weight, generating feature (3).

We find these distinctive electrodynamics are a generic feature of Fe(Te,Se) superconductivity, with qualitatively similar behavior observed in samples grown by different groups, grown via different techniques, and grown to different (Te,Se) concentrations, in both heterostructures and bare films (See SI Section 2). Likewise, previous terahertz studies \cite{Pimenov2013,Isoyama2021} of Fe(Te,Se) compounds can be understood within this framework (See SI Section 3). While similar anomalous electrodynamics featuring a sharp low-frequency absorption peak in both cuprates \cite{Corson1999,Bilbro2011,Corson2000} and disordered conventional superconductors \cite{Liu2013,Mondal2013,Swanson2014,ShermanNatPhys2015} indicate that similar phenomenology can emerge across diverse classes of superconductors, the underlying origins vary from system to system. Understanding these origins can therefore reveal key details of the fascinating superconducting state in Fe(Te,Se). 

\section{Fluctuating Superconductivity}\label{sec_DisorderedSC}

The anomalous electrodynamics of Fe(Te,Se) reported above result from strong, persistent fluctuations of the superconducting state. Such fluctuations correspond to the breakdown of phase coherence over long spatio-temporal scales, made possible by a low-density superfluid \cite{Corson2000}. The resulting phase fluctuations shift spectral weight from the condensate $\delta$-function to finite energy \cite{Orenstein2003,BasovTimusk2005}, representing a collective scattering of Cooper pairs driven between locally-coherent condensates. The anomalous fluid thus describes the dissipation due to superconducting fluctuations below $T_c$ and hence grows both with the strength of the fluctuations and with the weight of the total superfluid condensate. The proportionality $\kappa = k_a/k_s$ therefore characterizes the strength of the fluctuations. The spike in $\kappa$ just below $T_c$ for Fe(Te,Se) (see Fig. \ref{fig:Fig2}.e.) is consistent with strong fluctuations ordinarily present just below the superconducting transition when the condensate is quite weak. What is remarkable, however, is the quite large value of $\kappa$ even at the lowest temperatures, indicating the persistence of these strong superconducting fluctuations in Fe(Te,Se) even upon approaching the zero-temperature limit. 

\begin{figure*}
    \centering
    \includegraphics[width=\textwidth]{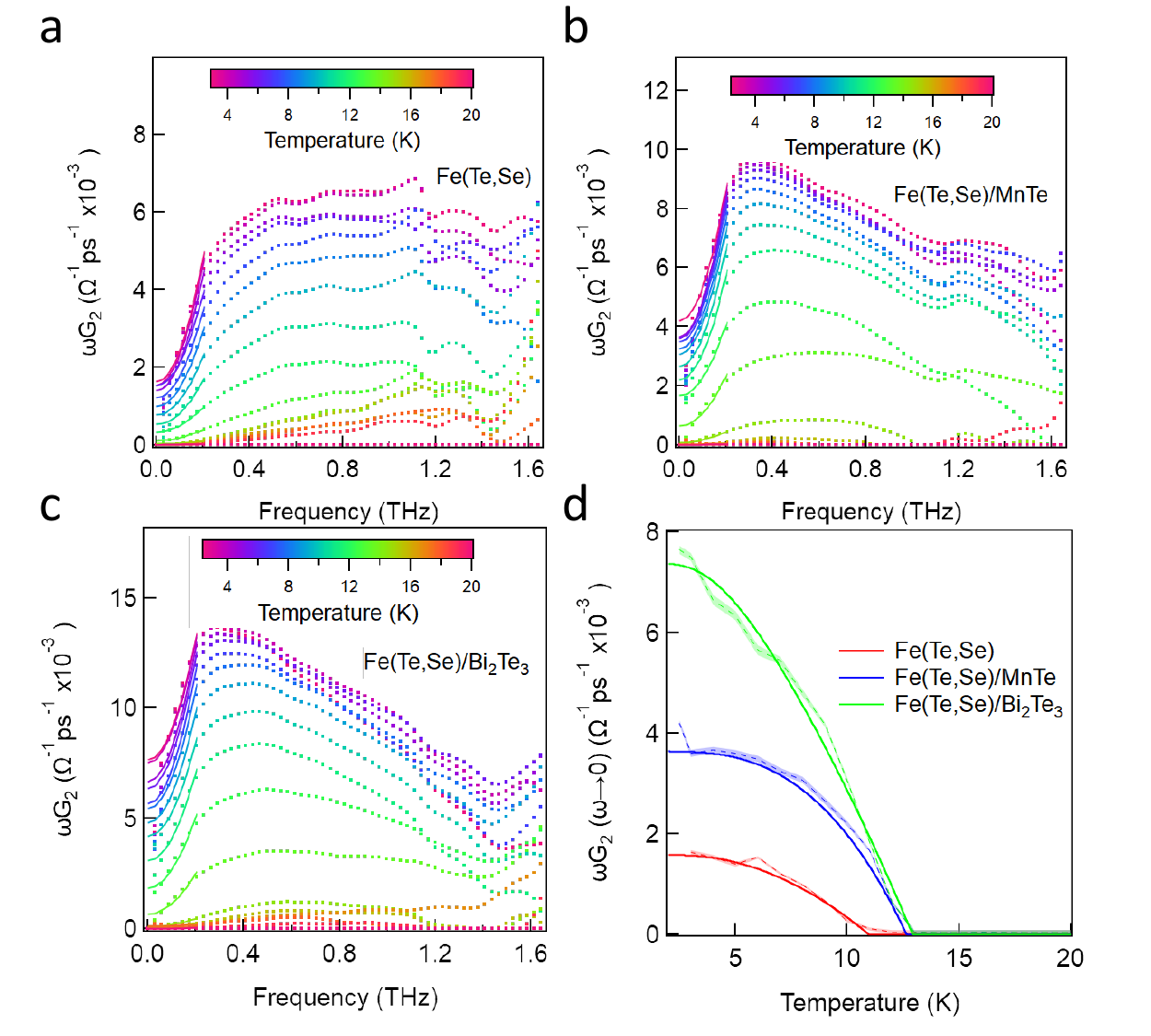}
    \caption{\textbf{Collapse of phase stiffness.} \textbf{a-c.} Phase stiffness of Fe(Te,Se), Fe(Te,Se)/MnTe, and Fe(Te,Se)/Bi$_2$Se$_3$ samples, respectively, with the 20 K phase stiffness subtracted away to eliminate common background contributions and anomalies. The low frequency limit is fitted with a quadratic function to extract the superfluid weight at zero frequency. \textbf{d.} Temperature dependence of superfluid weight at zero frequency (dashed line), fitted with the BCS expectation (solid line) \cite{AmbegaokarBaratoff1963,Stensberg2023} Uncertainty in fitted values are given by the shaded region.}
    \label{fig:Fig3}
\end{figure*}

This picture of fluctuating superconductivity is confirmed by the hallmark electrodynamic signature of collapsing phase stiffness at low frequency, which corresponds to the breakdown of phase coherence over long spatio-temporal scales. The phase stiffness $T_{\Theta}$ is given by
\cite{Corson1999,Bilbro2011}
\begin{equation}\label{EqnPhaseStiff}
    k_B T_{\Theta} = \hbar \omega \sigma_2 / \sigma_Q 
\end{equation}
\noindent where $k_B$ is the Boltzmann constant, $\Tilde{\sigma} = \sigma_1 + i\sigma_2 = \Tilde{G}/t$ is the complex conductivity of a thin film, and $\sigma_Q$ is the quantum conductivity. The quantum conductivity $\sigma_Q = e^2/\hbar d$ is determined by a characteristic length scale $d$, which is set by the coherence length $\xi$ of the superconductivity in Fe(Te,Se) \cite{EmeryKivelson1995}.  As the thickness $t$ of the superconducting layer can be ambiguous in heterostructures due to proximity effects, we first estimate the qualitative behavior of the phase stiffness from $T_{\Theta} \propto \omega G_2$. As shown in Fig. \ref{fig:Fig3}a-c, the phase stiffness drops rapidly at low frequency, but does not go fully to zero. According to the anomalous fluid model, the low-frequency limit of the phase stiffness should be given by $\omega G_2(\omega \to 0) \approx k_s + k_a\tau_a^2 \omega^2 $ (See SI Section 4). The model provides a good fit to the low-frequency data in each sample, shown by the solid lines in Fig. \ref{fig:Fig3}.a-c, confirming both a collapse of phase stiffness at low frequency and a weak but nontrivial superfluid in the DC limit, where the fitted values of $k_s$ are shown in Fig. \ref{fig:Fig3}.d. This collapse of phase stiffness at low frequency confirms that the superfluid struggles to enforce phase coherence throughout the superconducting regime.

This weak phase coherence results in the phase stiffness $T_{\Theta}$ of Fe(Te,Se) being comparable to its coherence temperature $T_c$. Using a coherence length of $\xi = 5$ nm \cite{Kasahara2016,Kasahara2014,Farrar2020} and Fe(Te,Se) film thickness of $t = 20$ nm, $T_{\Theta}$ can be calculated from Equation \ref{EqnPhaseStiff} using the low temperature data. This yields $T_{\Theta}(\omega) \approx 50$ K at higher frequencies, falling to $T_{\Theta}(\omega \to 0) \approx 12.8$ K in the DC limit (see \ref{fig:FigExtX_Ttheta}). The collapse of the phase stiffness in the low-frequency limit thus yields a ratio of $T_{\Theta}/T_c \approx 1$. This would imply that thermal fluctuations at $T_c$ should be sufficient to break the phase coherence even if the superconducting gap were to remain fully open at $T_c$. As suggested by Emery and Kivelson \cite{EmeryKivelson1995}, this ratio ratio of $T_{\Theta}/T_c \approx 1$ could indicate that superconductivity emerges in Fe(Te,Se) not because of a gap opening at that temperature, but rather because thermal fluctuations are sufficiently reduced to permit the establishment of phase coherence between preformed Cooper pairs.

Despite satisfying the Emery-Kivelson criteria for a phase-coherence-mediated superconducting transition, however, the characteristics of the remnant uncollapsed phase stiffness are nevertheless remarkably consistent with the opening of a superconducting gap on the hole band at $T_c$. As noted above, the quadratic fittings of $\omega G_2(\omega \to 0)$ in Fig. \ref{fig:Fig3}.a-c give the superfluid condensate weights $k_s$ as a function of temperature, which are shown in Fig. \ref{fig:Fig3}.d. The temperature dependence is remarkably well described by the usual Ambegaokar-Baratoff relation for the superfluid spectral weight $W(T) = \lim_{\omega \to 0} \omega G_2(\omega,T)$ of a BCS-like superconductor, given by \cite{AmbegaokarBaratoff1963,Stensberg2023} 
 \begin{equation}
    W(T) = W(0)\frac{\Delta(T)}{\Delta(0)}\tanh{\left(\frac{\Delta(T)}{2k_BT}\right)} 
 \end{equation}
\noindent where the temperature dependence of the superconducting gap is given by the BCS expectation $\Delta(T) \approx \Delta(0)\tanh{(1.74\sqrt{T_c/T-1})}$. These fittings provide estimates for the ratio $2\Delta/k_B T_c$, which takes the value 3.52 in BCS theory. The fittings yield ratios in the range of $\sim 2.7-3.3$ for the three Fe(Te,Se) samples (See SI Section 4 for table of fitted values). These ratios strongly disagree with previous estimates of $\sim~ 7-8$ \cite{Kasahara2016}, which were comparable to both cuprates \cite{Keimer2015}
and disordered conventional superconductors \cite{Sacepe2020}. This large disagreement is readily explained: previous ratio estimates were calculated with the larger superconducting gap $2\Delta_e \approx 8-10$ meV of the electron band; however, our fittings give effective gap estimates of $2\Delta \approx 3-4$ meV that align closely with the smaller superconducting gap $2\Delta_h \approx 3-5$ meV of the hole band \cite{Zhang2018,Kasahara2016}. Despite the multiband character of Fe(Te,Se), the terahertz conductivity would suggest the electrodynamics of the superconducting state exhibit a $T_c$ and effective $2\Delta(T)$ consistent with a BCS-like weak coupling state initiated by the hole band. 

While the initiation of phase coherence below $T_c$ appears consistent with the opening of an weak-coupling effective gap, evidence of incoherent preformed Cooper pairing also appears above $T_c$, fulfilling the Emery-Kivelson expectation and more consistent with effective strong-coupling superconductivity \cite{EmeryKivelson1995}. A close analysis reveals a finite phase stiffness above the nominal $T_c$, especially visible at higher frequencies in Fig. \ref{fig:Fig3}.a-c. The continuation of this phase stiffness across $T_c$ (see \ref{fig:FigExtX_wG2_temps}) indicates the persistence of Cooper pairs above $T_c$, despite the lack of a coherent superconducting state. Such superconducting fluctuations above the nominal $T_c$ due to incoherent preformed Cooper pairs are similar to previous terahertz observations in cuprates \cite{Corson1999,Bilbro2011,Corson2000}. However, in contrast to the smooth increase in phase stiffness alongside the progressive emergence of superconductivity in the Emery-Kivelson picture \cite{EmeryKivelson1995}, the Fe(Te,Se) samples exhibit an abrupt increase in phase stiffness below $T_c$. This rapid increase again suggests that, while the population of preformed Cooper pairs is indeed increasing, it is the effective gap opening that initiates phase coherence in Fe(Te,Se). Altogether, Fe(Te,Se) hosts remarkably persistent superconducting fluctuations both above and below $T_c$.

\section{Scaling Analysis of Fluctuations and Planckian Dissipation}\label{sec_SIT}

\begin{figure*}
    \centering
    \includegraphics[width=\textwidth]{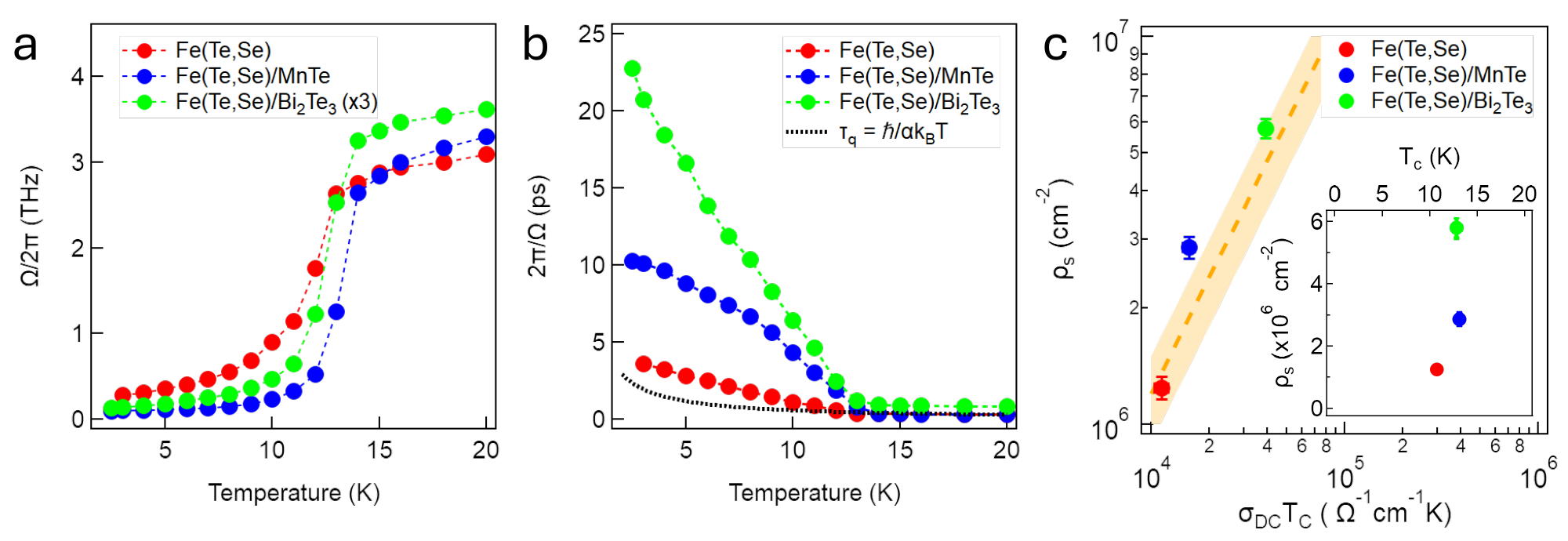}
    \caption{\textbf{Universal Scaling Characteristics.} \textbf{a.} Temperature-dependent fluctuation frequency for the Fe(Te,Se), Fe(Te,Se)/MnTe, and Fe(Te,Se)/Bi$_2$Se$_3$ samples derived from the rescaling procedure. \textbf{b.} Corresponding fluctuation timescale compared to the scattering time $\tau_q$ of Planckian dissipation, using $\alpha = 1.38$. \textbf{c.} Universal scaling parameters of the samples compared in to Homes's scaling and Uemura's scaling (inset). The dashed line indicates the usual proportionality in Homes scaling, $\rho_s = (120 \pm 25) \sigma_{DC} T_c$ \cite{HomesNature2004}.}
    \label{fig:Fig4}
\end{figure*}

The intimate link between fluctuation and dissipation enables universal scaling analysis\cite{Schmidt1968, Skocpol1975, FisherFisherHuse1991, DorseyPRB1991} of the superconducting fluctuations in Fe(Te,Se).  As demonstrated in disordered conventional superconductors \cite{Liu2013,Mondal2013} and cuprates \cite{Corson1999,Bilbro2011}, the complex conductance in the fluctuation regime can be rescaled by a temperature-dependent fluctuation frequency $\Omega(T)$ to yield a scaling function $\Tilde{S}(\omega/\Omega(T)) = S_1 + iS_2$ consisting of an amplitude $|S|$ and phase $\Phi = \textrm{tan}^{-1}(S_2/S_1)$ that converge upon universal curves. The rescaling of the conductivity represents the noise spectrum \cite{HoulrikPRB1994,SchottkyPR1926} of the system and is accomplished via the equation \cite{Corson1999,Bilbro2011}
\begin{equation}\label{EqnRescale}
    \Tilde{S}(\omega/\Omega(T)) = \left(\frac{\hbar \Omega(T)}{k_B T_{\Theta}^0(T)}\right)\frac{\Tilde{\sigma}(\omega,T)}{\sigma_Q}
\end{equation}
\noindent where $T_{\Theta}^0(T)$ is the bare phase stiffness determined from the data. (See SI section 5 for rescaling procedure). Rescaling the conductivity according to Eqn \ref{EqnRescale} (see \ref{fig:FigExtX_Scaling}) yields the temperature-dependent fluctuation frequency $\Omega(T)$ for the three samples shown in Fig. \ref{fig:Fig4}.a. $\Omega(T)$ is expected to rapidly fall to zero as phase coherence develops at $T_c$; that is, the fluctuation timescale $2\pi/\Omega(T)$ diverges. Yet while $\Omega(T)$ in Fe(Te,Se) rapidly drops across $T_c$, as shown in Fig. \ref{fig:Fig4}.a, the fluctuation frequency remains finite down to the lowest temperature, confirming the persistence of superconducting fluctuations. This is yet more apparent in Fig. \ref{fig:Fig4}.b, where the fluctuation time fails to diverge at any temperature, pointing to quantum fluctuations being the ultimate origin of the persistent superconducting fluctuations in Fe(Te,Se) far below $T_c$.

The fluctuation timescale $2\pi/\Omega(T)$ undergoes a sharp hinge point in the slope at $T_c$, as seen in Fig. \ref{fig:Fig4}.b. Approaching $T_c$ from below, it can be seen that this hinge point roughly coincides with $2\pi/\Omega(T)$ reaching the Planckian timescale $\tau_q = \hbar/\alpha k_B T$, where $\alpha$ is a scale factor of near-unity \cite{ZaanenNature2004,HartnollMackenziePRMP2022,PhillipsScience2022}. As illustrated in Fig. \ref{fig:Fig4}.b, the experimental scale factor for Fe(Te,Se) is $\alpha = 1.38 \pm 0.05$. That is, Fe(Te,Se) undergoes a fairly abrupt transition from strongly fluctuating superconductivity below $T_c$ to a normal state undergoing Planckian dissipation above $T_c$. This implies that the establishment of superconducting coherence, and hence $T_c$, coincides with the moment when the timescale of superconducting fluctuations exceeds the Planckian timescale. 

Planckian dissipation is associated with strange metal states and quantum critical fluctuations, exhibiting linear-in-$T$ resistivity in segments of many unconventional superconductor phase diagrams \cite{Keimer2015,PhillipsScience2022,HartnollMackenziePRMP2022}. Similar linear-in-$T$ behavior in the precursor state above $T_c$ is observed for the Fe(Te,Se) samples measured here (see Fig. \ref{fig:Fig1}.j). The Planckian dissipation of these precursor states generally causes \cite{ZaanenNature2004} the superfluid density $\rho_s$ below $T_c$ to obey the Homes scaling relation\cite{HomesNature2004}, $\rho_s \propto \sigma_{DC} T_c$. This contrasts with the more restrictive Uemura scaling relation\cite{UemuraPRL1989}, $\rho_s \propto T_c$, which obtains primarily when a pronounced pseudogap phase intervenes between the strange metal and superconducting phases, as in underdoped cuprates \cite{Keimer2015} and highly disordered conventional superconductors \cite{Sacepe2020,ImryPRL2012, Charpentier2025}. As illustrated in Fig. \ref{fig:Fig4}.c and inset, respectively, the three Fe(Te,Se) samples in this study do indeed obey the Homes scaling while failing the Uemura scaling (see SI Section 6), confirming the consistency of Planckian-dissipative behaviors above $T_c$ with the observed characteristics of the samples below $T_c$.  With Planckian dissipation previously reported \cite{Jiang2023} in the FeSe parent compound, it appears likely that the states above and below $T_c$ are linked by Planckian dissipation across extended regions \cite{PhillipsScience2022,HartnollMackenziePRMP2022} of the Fe(Te,Se) phase diagram.

\section{Discussion}\label{sec_Discussion}

The transition from a Planckian-dissipative precursor state to persistent fluctuating superconductivity in Fe(Te,Se) appears to be inextricably linked to underlying fluctuations. The multiband nature of Fe(Te,Se) seems to produce a kind of mixed-limit superconductivity with both weak-coupling and strong-coupling characteristics, as well as a large residual uncondensed fluid. The resulting state is dominated by strong fluctuations throughout the superconducting regime, which fail to diminish at the lowest temperatures, and which converge upon Planckian dissipation above $T_c$. This points to extended quantum fluctuations as the origin of both the persistent fluctuating superconductivity and Planckian dissipation in Fe(Te,Se). Understanding these underlying fluctuations would be greatly aided by spatially- and temporally-resolved probes of both superconducting and Planckian fluctuations \cite{KremenNatPhys2018,BoschiniNatMater2018,ChenScience2023}, with broad applicability to emerging classes of unconventional superconductors where quantum fluctuations may similarly weaken the coherence of the superconducting state.

\bibliography{FTS_dSC}

\section{Methods}\label{sec11}

\textbf{Sample Growth}
 
Samples of FeTe$_{0.70}$Se$_{0.30}$ are grown by molecular beam epitaxy to a thickness of $t = 20$ nm with substrates transparent to THz. In the case of the thin film Fe(Te,Se) sample, the Fe(Te,Se) is grown directly on a CaF$_2$ substrate, which has good lattice matching with Fe(Te,Se). For the heterostructures of Fe(Te,Se)/MnTe and Fe(Te,Se)/Bi$_2$Te$_3$, 10 nm layers of MnTe and Bi$_2$Te$_3$ are grown on Al$_2$O$_3$ substrates, with hybrid symmetry epitaxy \cite{Yao2022,Yao2021} being used to subsequently grow the square-lattice Fe(Te,Se) atop the hexagonal-lattice MnTe or Bi$_2$Te$_3$, respectively. In both cases, the fortuitous close-matching of the rotated lattices permits highly quality Fe(Te,Se) growth despite the difference in lattice geometry without the need for a buffer layer. The samples are measured by DC transport in van der Pauw geometry to ensure a clear transition to the superconducting state at low temperature.

\textbf{Time-domain Terahertz Spectroscopy}

A femtosecond pulsed laser (780 nm; 82 fs pulse duration; 10 mW power; 80 MHz repetition rate) is split into two beams, which irradiate a pair of photoconductive antennas (PCA): an emitter and a detector. The illuminated emitter PCA produces a quasi-single cycle THz pulse that is transmitted through the sample (or reference substrate) to the detector PCA. When illuminated by the femtosecond pulse, the detector PCA produces a photocurrent proportional to the electric field of the THz pulse at that time. Scanning the time delay between the THz and femtosecond pulses as the detector PCA therefore coherently maps the electric field profile of the THz pulse in the time, the Fourier transform of which yields the spectrum of the THz pulse. In order to avoid spectral anomalies due to water vapor, the entire THz spectrometer is enclosed in a box that is continuously purged with dry air.

As TDTS is a fully coherent measurement of amplitude and phase, the complex transmission $\Tilde{T}$ of a sample can be extracted by dividing the sample spectrum by the substrate spectrum. The complex conductance $\Tilde{G}$ for the sample can then be calculated from $\Tilde{T}$ according to the thin film relation
\begin{equation}
    \Tilde{G}(\omega) = \frac{n+1}{Z_0} \left( \frac{1}{\Tilde{T}(\omega)}e^{i\omega\Delta L(n-1)/c}-1 \right)
\end{equation}
\noindent where $n$ is the substrate index of refraction, $Z_0$ is the vacuum impedance, $c$ is the speed of light, and $\Delta L$ is the difference between the thickness of the reference substrate and the thickness of the substrate on which the sample is grown. The exponential term accounts for the phase mismatch produced by $\Delta L$. For the thin film limit, it is generally considered that the complex conductance $\Tilde{\sigma} = \Tilde{G}/t$, where $t$ is the sample thickness. TDTS data are initially reported in $\Tilde{G}$ to avoid hidden assumptions about the homogeneity of the conducting state throughout the thickness of the sample, which may be ambiguous for heterostructures. Values of $t$ are subsequently specified for all calculations employing $\Tilde{\sigma}$. (See SI Section 1 for further details)

\section{Acknowledgments}

We thank Peter Armitage for helpful discussions. J.S. was partially supported by the Gordon and Betty Moore Foundation’s EPiQS Initiative under the grant GBMF9212 to L.W.  and the Dissertation Completion Fellowship at the University of Pennsylvania. P.M.T acknowledges partial support from the Croucher Foundation. C.L.K was supported by a Simons Investigator grant from the Simons Foundation. I.T. are supported by AFOSR FA9550-14-10332. X.Y. and S.O. were sponsored by the Army Research Office and was accomplished under Grant Number W911NF-20-2-0166. H.Y., C.-Y.L. and I.T. were supported by AFOSR FA9550-14-10332 and NIST Grant No. 60NANB19D027 with partial support from Laboratory of Physical Sciences and NSF DMR-2329087. A.-H. C and M. B. were supported by the U. S. Department of Energy, Office of Science, Basic Energy Sciences, Materials Sciences and Engineering Division. L.W. acknowledges support from the Army Research Office under Grant Number  W911NF-25-2-0016 and the Sloan Foundation under the award FG-2025-25036. 

\setcounter{figure}{0}
\renewcommand{\figurename}{}
\renewcommand{\thefigure}{Ext. Data Fig. \arabic{figure}}

\begin{figure*}
    \centering
    \includegraphics[width=\textwidth]{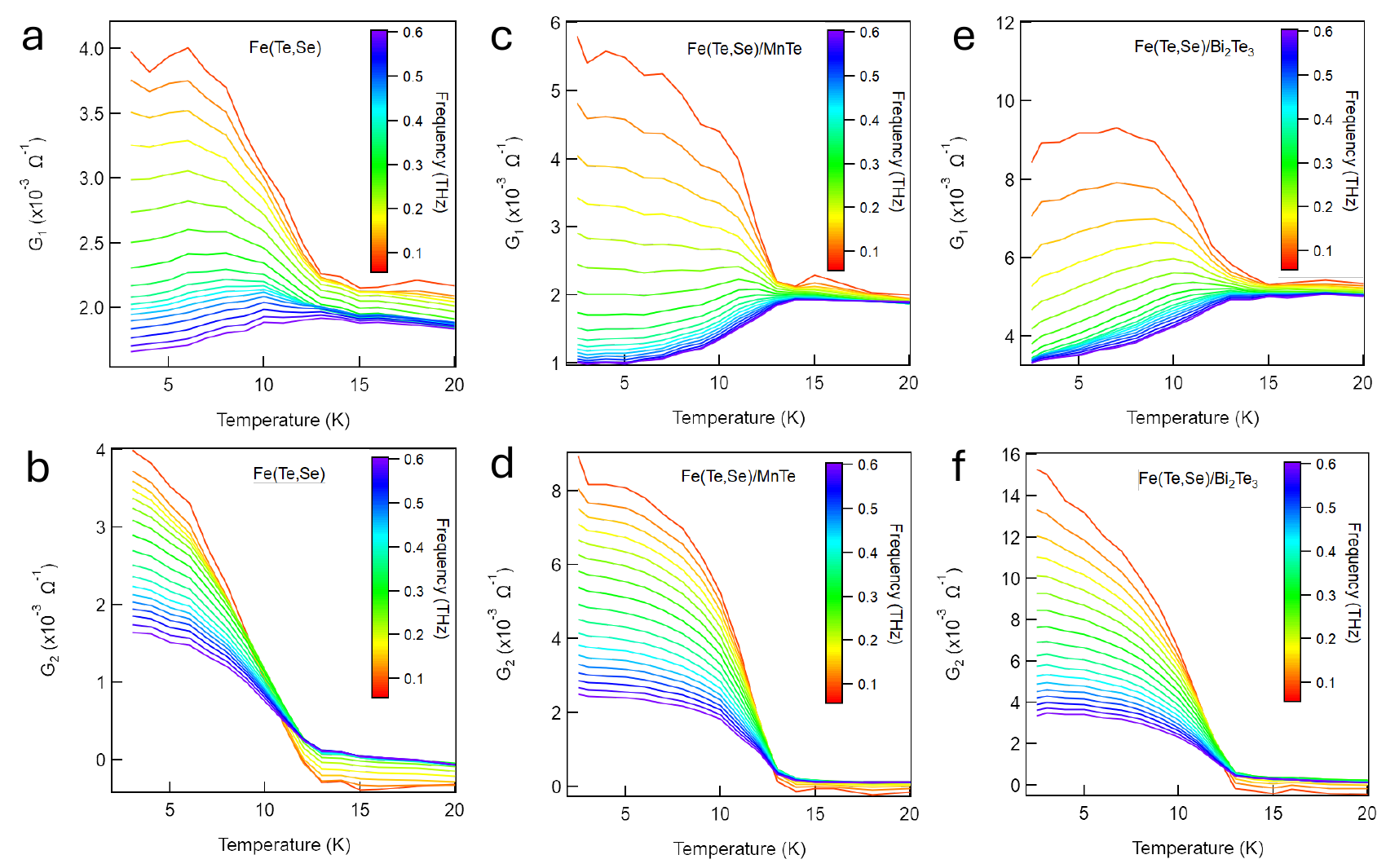}
    \caption{\textbf{Temperature Dependence of Conductance.} The temperature dependence of $G_1$ and $G_2$ for \textbf{a,b} Fe(Te,Se), \textbf{c,d} Fe(Te,Se)/MnTe, and \textbf{e,f} Fe(Te,Se)/Bi$_2$Te$_3$ for frequencies spanning 0.09 - 0.60 THz. The three key electrodynamic features of Fe(Te,Se) below $T_c$ can be clearly distinguished in the low frequency and high frequency responses, and their persistence to the lowest available temperatures is apparent.}
    \label{fig:FigExtX_G1G2_temps}
\end{figure*}

\begin{figure*}
    \centering
    \includegraphics[width=\textwidth]{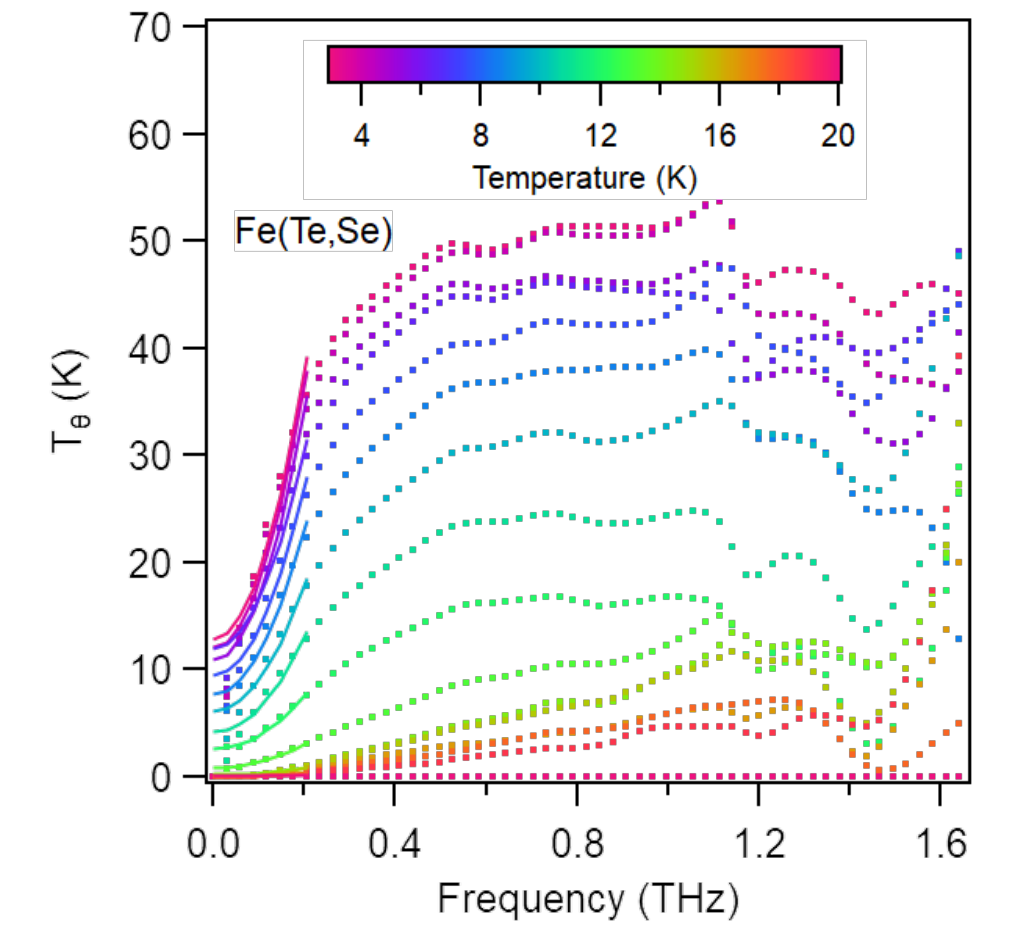}
    \caption{\textbf{Phase Stiffness Temperature.} Phase stiffness temperature can be accurately estimated from Equation \ref{EqnPhaseStiff} for Fe(Te,Se) using the sample thickness $t = 20$ nm and an estimate of the coherence length $\xi = 5$ nm \cite{Kasahara2016,Kasahara2014,Farrar2020}. In the low frequency limit, the data is fitted by a quadratic dependence predicted by the anomalous fluid model.}
    \label{fig:FigExtX_Ttheta}
\end{figure*}

\begin{figure*}
    \centering
    \includegraphics[width=\textwidth]{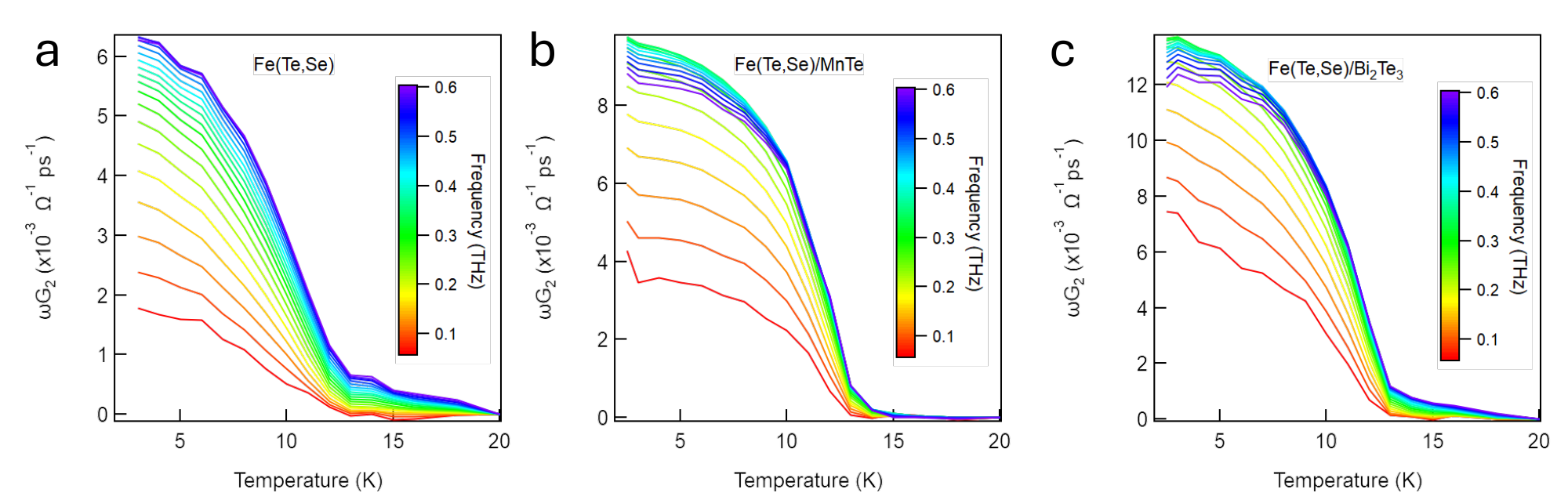}
    \caption{\textbf{Temperature Dependence of Phase Stiffness.} The temperature dependence of $\omega G_2$ proportional to the phase stiffness for \textbf{a.} Fe(Te,Se), \textbf{b.} Fe(Te,Se)/MnTe, and \textbf{c.} Fe(Te,Se)/Bi$_2$Te$_3$, relative to 20 K. In all three samples, the distinctive growth of higher-frequency phase stiffness increases above the nominal superconducting transition temperature of 13 K suggests the presence of preformed Cooper pairs.}
    \label{fig:FigExtX_wG2_temps}
\end{figure*}

\begin{figure*}
    \centering
    \includegraphics[width=\textwidth]{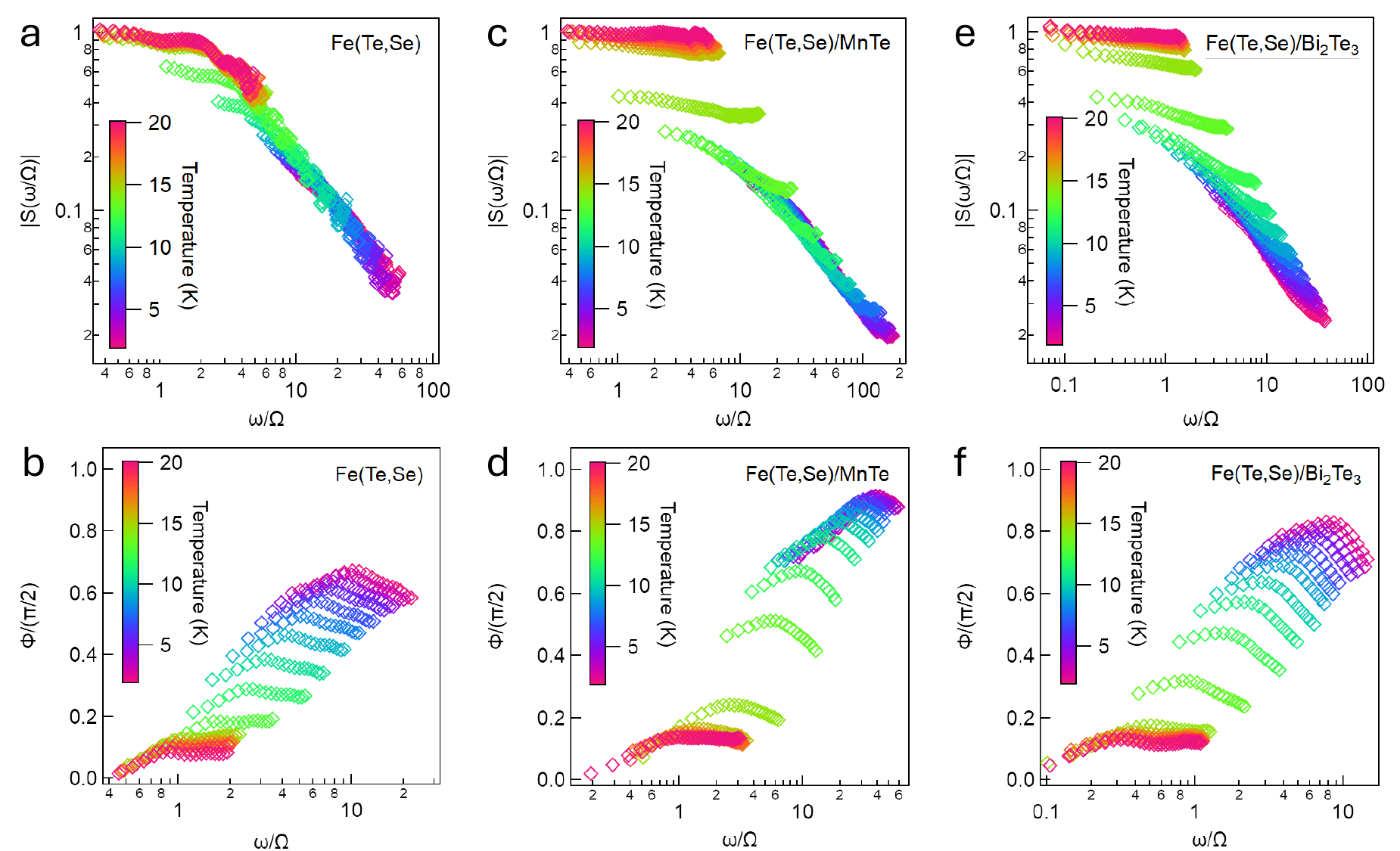}
    \caption{\textbf{Scaling Analysis.} Rescaled amplitude $|S(\omega/\Omega(T)|$ and phase $\Phi = \tan^{-1}(S_2/S_1)$, respectively, of the terahertz data for \textbf{a,b.} Fe(Te,Se), \textbf{c,d.} Fe(Te,Se)/MnTe, and \textbf{e,f.} Fe(Te,Se)/Bi$_2$Te$_3$. In each case, the rescaled data converges upon the same universal curves for fluctuating superconductivity, despite spanning the whole of the superconducting regime for Fe(Te,Se). Discontinuities in the curves--jumps between temperatures--results from relatively large temperature steps (1 K) in regions of rapidly changing behavior around $T_c$. Deviations from total collapse of the data onto perfectly smooth curves results from residual contributions to the conductivity that are inherently not captured by the fluctuation model.}
    \label{fig:FigExtX_Scaling}
\end{figure*}

\end{document}